# Self-energy corrected band-gap tuning induced by strain in the hexagonal boron phosphide monolayer

Jose Mario Galicia-Hernandez [a*], J. Guerrero-Sanchez [a], R. Ponce-Perez [a], H. N. Fernandez-Escamilla [a], Gregorio H. Cocoletzi [b], and Noboru Takeuchi [a].

[a] *Centro de Nanociencias y Nanotecnología, Universidad Nacional Autónoma de México, Carretera Tijuana-Ensenada km 107, Apdo., 22860, Ensenada, B.C., México*

[b] *Instituto de Fisica "Ing. Luis Rivera Terrazas", Benemerita Universidad Autonoma de Puebla, Av. San Claudio & Blvd. 18 Sur, Ciudad Universitaria, Colonia San Manuel, C.P. 72570, Puebla, Puebla, Mexico.*

\* Corresponding author: josemariogahe@gmail.com

*Abstract:* We performed a first-principles study of the electronic behavior of a 2D hexagonal boron phosphide monolayer (2D-h-BP). The system was deformed isotropically by applying a simultaneous tensile strain along the *a* and *b* crystal axes. We analyzed the band-gap evolution as function of the deformation percentage, ranging from 1% to 8%. Results show that the system behaves as a direct band-gap semiconductor, with the valence band maximum and conduction band minimum located at the K point (1/3, 1/3, 0) of the Brillouin zone. This behavior is unchanged despite the strain application. The band gap underestimation, as computed within the standard DFT, was corrected by applying the $G_0W_0$ approach. Trends in the band-gap behavior are the same within both approaches: for low deformation percentages, the band-gap grows linearly with a small slope, and at higher values, it grows very slow with a tendency to achieve saturation. Therefore, the band-gap is less sensitive to tensile strain for deformations near 8%. The origin of this band gap behavior is explained in terms of the projected density of states and charge densities, and it can be attributed to Coulomb interactions, and charge redistributions due to the applied tensile strain. To study the carrier mobility, we computed the electron and hole effective masses, finding high mobility for both carriers. Finally, the stability analysis of each strained system includes the calculation of phonon spectra, to assure the dynamical stability, the computation of elastic constants to evaluate the mechanical stability, and cohesive energies for exploring the thermodynamical stability. Results indicate that the boron phosphide monolayer is stable under the calculated tensile strains.

*Keywords:* Density functional theory, GW approach, strain-induced band gap tuning, hexagonal boron phosphide monolayer.

## 1. Introduction

The rise of two-dimensional (2D) materials has brought a new field of study. Here, the star is graphene, because of its extensive list of exciting properties: it is the strongest 2D material to date, it is flexible, possesses massless charge carriers, and it has a high electronic conductivity, among others [1-5]. However, it is a zero band-gap material, precluding its direct application into electronic devices. For this reason, several ways to engineer its band-gap have been recently investigated [6, 7]. On the other hand, the scientific community continues to look for other 2D materials with new and novel properties. In particular, it is known metals belonging to groups 13 and 15 can form low dimensional systems on semiconductor systems [8-11]. Therefore, it is possible that these elements can also form stable free standing 2D systems. For example, phosphorene is another 2D material under intensive research. It is an intrinsic direct-transition semiconductor in which the band-gap can be tuned by strain. Within this technique, the system can experience a change in electronic behavior, turning from a direct to indirect semiconductor [12-14]. Other possibilities are the stacking of several layers [15, 16], and using an external electric field [17]. One of the most critical limitations of phosphorene is its high activity towards oxygen: in ambient conditions, it tends to oxidize, precluding its use in electronic devices operating at normal conditions [18, 19]. Another interesting material is the wide band-gap hexagonal boron nitride monolayer (2D h-BN). Its electronic character assures light emission at ultraviolet and deep ultraviolet spectra [20]. The 2D h-BN monolayer is a dielectric that may be used in heterostructures with other 2D materials such as graphene [21-23]. Although 2D h-BN monolayer has several interesting properties and applications, its large band-gap precludes its use in visible light range devices.

Another 2D material with similar characteristics is the hexagonal boron phosphide monolayer (2D-h-BP). It has a shorter band-gap, as needed for applications in the visible region of the electromagnetic spectrum, and with high carrier mobility. This is a 2D material in which the B and P atoms are covalently bonded with s$p^2$ hybridization, forming a honeycomb lattice structure similar to graphene with a flat atomic structure. 2D-h-BP possesses a direct band gap with a moderate value, high mechanical stability with a Young's modulus equal to 139 N/m [24], and high carrier mobilities: hole mobilities are $1.37 \times 10^4$ and $2.61 \times 10^4$ cm$^2$/Vs along zigzag and armchair directions respectively, and for electrons, $5.00 \times 10^4$ and $6.88 \times 10^4$ cm$^2$/Vs along the same directions [25]. These values are remarkably larger than those of MoS$_2$ and phosphorene. 2D-h-BP possesses a strong optical absorption from 1.37 eV to 4 eV [26] which makes it a promising material for electronic and optoelectronic applications to develop high-performance devices such as field-effect transistors.

The stability of 2D h-BP was predicted using first principles calculations [25-31]. However, up to now, 2D-h-BP has not been successfully synthesized by experiments. Actually, the synthesis of BP in bulk is also challenging [32]. However, previous experimental works suggest that the synthesis of 2D-h-BP could be achieved in the near future: the growth of single crystal layers of B-P compounds on Si substrates has been achieved [33]. In the same way, a boron phosphide film was grown on the (111) Si substrate using the atmospheric-pressure MOCVD method [34]. Besides, the growth of BP films on SOI, sapphire, and silicon carbide substrates using CVD has been achieved [35, 36]. Other experimental work reveals the heteroepitaxial growth of high-quality BP films on a superior aluminum nitride (0001)/sapphire substrate by CVD [37]. Finally, a theoretical study based on density functional theory raises the possibility of obtaining 2D-h-BP by exfoliation of the BP (111) surface [38]. Potential applications of 2D-h-BP have been predicted in many fields. Few examples are its use in batteries [39, 40], in catalysis [39], or in optoelectronic and spintronics devices [26, 29, 41-43].

The band gap is a key intrinsic property to understand the electronic behavior of a material. Its value and type (direct or indirect) determine the number of potential applications, that can be widened if it can be tuned up to a desired value. Therefore, theoretical studies have been carried out in the case of 2D-h-BP. For example, the band gap of the monolayer can be altered if dangling bonds are fully saturated by hydrogen atoms [44]. However, for certain applications this technique is not convenient since a very large band gap (4.80 eV) is obtained, and the system changes its electronic behavior to become an indirect semiconductor (a transformation that is not good for optical applications). Besides, a structural variation in the monolayer takes place after the addition of H atoms, with a change from a planar structure into a buckled geometry. An electric field has also been used to tune the band gap of 2D-h-BP. By this technique, a dramatic change in electronic properties is obtained, reaching a pseudo-Lifshitz phase transition in the material [45]. Although the electron mobility is increased, the band gap is reduced in such a way that the band structure adopts a graphene-like configuration, which limits the range of potential applications in electronic devices (such as transistors) where a high on/off ratio is required. Theoretical studies have shown that the electronic properties of 2D-h-BP, especially the band gap, can be tuned up by functionalization with either Br [46], or Cl [30] atoms. Although this technique is a good alternative for band gap tuning, it can lead to the creation of magnetic moments in the monolayer. Besides, the presence of adsorbed atoms may not be convenient, since they can be seen as impurities in operando environments of certain electronic devices where the monolayer can be used. A similar situation happens by doping the monolayer with elements of groups III, IV and V [47], or by inducing vacancies in the monolayer [48].

On the other hand, it is possible to alter the electronic structure of semiconductors by the application of external pressure. This is especially true in 2D systems, due to their high elastic limits in comparison with their bulk counterparts. The change in interatomic distances, as well as the relative positions of the atoms lead to modifications in the electronic band structures and densities of states of the systems under external pressure. In a certain range of deformations, the structural properties are not deeply changed, and the systems preserve the same kind of structure (especially when tensile strain is applied). For small deformations, the symmetry is also preserved. One of the most important advantages of strain-based band gap engineering is that the chemical properties and compositions are kept untouched, since the method is purely mechanic. There are some reports devoted to the study of electronic properties of 2D-h-BP under strain [31, 49, 50,]. However, they are focused on presenting a description of results and tendencies. A deep study that provides an explanation of the behavior of the band gap of 2D-h-BP under strain is still lack in the field. For this reason, in this paper, we have included a clear explanation related with the origin of the band gap opening as a consequence of tensile strain, in terms of projected density of states and charge densities. We use quasiparticle self-energy corrected calculations, and compare our results with previous tight binding and HSE functional calculations for the band-gap. We point out the importance of using self-energy corrected calculations to get a deeper insight into the full description of 2D-h-BP electronic properties.

Finally, it is worth to mention that 2D-hBP has potential applications in flexible electronic devices. These devices are formed by a substrate, which is responsible of providing the mechanical flexibility, the functional material (as 2D-hBP or other 2D system), which provides the functionality of the device, and finally an encapsulation layer whose function is to protect the functional material [51]. In general, during the operation of flexible devices, the functional material (which has a high flexibility and strain sensitivity) is subjected to distortions, bending (in the so-called bending devices), and stretching deformations [51]. These deformations are the origin of the strains in the functional material of a device during its operation. 2D systems are ideal for flexible devices because they are able to resist high elastic strains due to strong covalent bonds of its constitutive atoms and its small cross-sectional area [52], in this way the device can operate without a damage in the functional material. The strain can also be induced in the 2D functional material if there is a mismatch of lattice constants between the substrate and the functional material [53], for this reason the substrates chosen for flexible devices has a high Young's modulus in order to effectively induce strain to 2D functional materials [53]. Finally, a strain can also be transmitted to the 2D functional material if there is a thermal-expansion mismatch, i.e., a thermal strain is induced because of the difference of thermal expansion coefficients between the substrate and the 2D system [52, 53]. Therefore, inducing strains in 2D systems in a flexible device is found to be convenient since the lattice structure of the system can be changed and manipulated by this mechanism leading to the possibility of tuning the electronic properties and phonon spectra of the system without losing its stability what makes possible the application of these systems in several devices such as FETs, flexible strain sensors and flexible photodetectors [53].

This paper is organized as follows: the computational details for all calculations based on density functional theory are given in section 2. In section 3, a detailed description of structural and electronic properties is provided, followed by the computation of effective masses to study the carrier mobilities. In the last part of section 3 we present the stability analysis for all systems under study. Finally, a summary of results and conclusions are presented in section 4.

## 2. Computational details

Our first-principles calculations were based on the periodic density functional theory as implemented in Vienna Ab initio Simulation Package (VASP) [54, 55]. The electron-ion interaction was treated with PBE-PAW pseudopotentials. The exchange-correlation energy is treated within the Generalized Gradient Approximation (GGA) [56] in the parametrization of Perdew-Burke-Ernzerhof [57]. The electron wave function was expanded using a plane-waves basis set with a kinetic energy cut-off of 680 eV. The convergence criteria for the difference in energy between two steps in the self-consistent loop was set to $1 \times 10^{-4}$ eV. The atomic positions were optimized using the conjugate gradient method, with the convergence criteria that the maximum value allowed for forces was set to $8 \times 10^{-4}$ eV/Å. The system was modeled with a $1 \times 1$ supercell. A vacuum gap of width 25 Å along the $z$ direction is included to avoid interaction between adjacent layers. The Brillouin zone sampling was done with a *k*-points mesh according with the Monkhorst-Pack scheme [58]. A $21 \times 21 \times 1$ mesh was assigned for ionic relaxation calculations as well as for computation of band structures, phonon spectra, computation of elastic constants, calculation of cohesive energies, density of states and charge densities, and a mesh of $12 \times 12 \times 1$ was used to carry out the corrections of band-gap within $G_0W_0$ approach [59]. Van der Waals interactions have not been included in the present work, since we are considering a single BP monolayer.

The effectives masses were calculated from the curvatures of the band structures ($E$ vs. $k$ function). To have a good description of the curvatures, we used a very dense set of points on each segment between two symmetry points (150 *k*-points). Since standard DFT provides trustworthy qualitative results of the shape of the bands, we used this level of approximation to calculate the curvatures of $E$ as function of $k$.

In general, the effective mass tensor of the electron/hole is obtained from the curvature of band structure by using the following expression:

$$[m^{*-1}]_{i,j} = \frac{1}{\hbar^2}\frac{\partial^2 E}{\partial k_i \partial k_j}\bigg|_{k_{ext.}} \quad (1),$$

where $[m^*]_{i,j}$ refers to each component of the effective mass tensor obtained from the second partial derivative of the $E$ $vs$ $k$, and $k_{ext.}$ refers to the *k*-point at which an extremum is located (maxima or minima). Note that $E$ $vs.$ $k$ is a three-variable function, since each *k*-point has three components. However, it is possible to make a simplification if we consider $E$ $vs.$ $k$ as a one-variable function, similar to what we usually do when we plot band structures along high symmetry paths. If this consideration is taken into account, then, $m^*$ is a scalar and the partial derivative the becomes a total derivative. In our case, it is convenient to consider the function $E$ vs. $|\vec{k}|$, where $|\vec{k}|$ is defined as the magnitude of the vector which goes from the Γ point (the center of Brillouin zone, $k_\Gamma(0,0,0)$) to a specific *k*-point in the first Brillouin zone $k_i(k_1, k_2, k_3)$. The coordinates of each point are defined with respect to the axes of the reciprocal lattice generated by the vectors $b_1$, $b_2$ and $b_3$ (in our case $b_1 = b_2$). The considered high symmetry points and their corresponding reduced coordinates are Γ (0, 0, 0), K (1/3, 1/3, 0) and M (1/2, 0, 0), and their respective paths are the K → Γ and K → M segments. The computed band structure considers the reduced coordinates of each *k*-point, so to be consistent with equation 1 we multiplied each reduced coordinate by the vector length $b_1$. As we are dealing with a 2D system, the $k_3$ coordinate is equal to zero and will be omitted. To compute the effective mass from equation 1 (it involves a second derivative), we fitted the band structure to a polynomial function of degree $9^{th}$ (which perfectly matches the real function in the vicinity of the point K). The curvature of the polynomial function was obtained by calculating its second derivative evaluated at the k-point value at which the extremum is located. By substituting this value in equation 1, we get the corresponding effective mass.

## 3. Results and discussion

In this section, the results are presented, in the first subsection, we describe the system under study and the structural properties, secondly, we discuss in detail the electronic properties, finally, in the last subsection we discuss the dynamical, mechanical and thermodynamic stability.

### 3.1 Structural properties

The hexagonal boron phosphide monolayer has been modeled as a 2D periodic system consisting of a $1 \times 1$ supercell (with two atoms per unit cell) and a corresponding vacuum region of 25 Å perpendicular to the *xy*-plane to avoid interaction between adjacent layers (see figure 1 for details). A full relaxation was performed, i.e., the atomic positions as well as lattice constants were optimized. At the end of the relaxation process, the initial symmetry condition is preserved, i.e., $a = b \neq c$, and $\alpha = \beta = 90°, \gamma = 120°$. The optimized structural parameters were: $a = b = 3.212$ Å, B – P bond length =1.854 Å, and the B – P bond angle =120°.

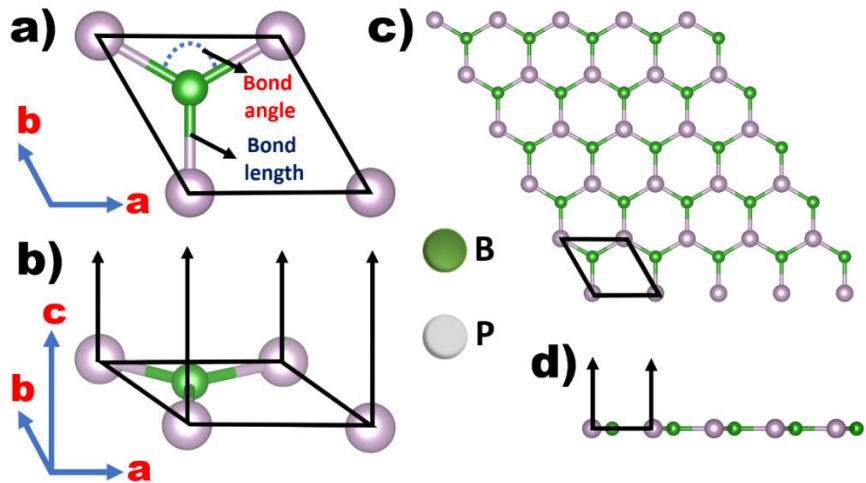

**Figure 1.** Structural model of the hexagonal boron phosphide monolayer: a) top view of the unit cell, b) side view of the unit cell indicating the vacuum region perpendicular to the plane, c) top view of the extended periodic system with multiple unit cells to show the honeycomb pattern, d) side view of the extended system to show that the monolayer is flat.

To construct the systems under tensile strain we expand the *a* and *b* lattice constants simultaneously and with the same proportion according with the expression:

$$\varepsilon_\% = \frac{a - a_0}{a_0} \times 100 \qquad (2),$$

where $\varepsilon_\%$ is the percentage of the lattice constant deformation due to the tensile strain, $a$ is the lattice constant after deformation, and $a_0$ is the lattice constant of the unstrained system. As the tensile strain is applied simultaneously along both lattice constants $a$ and $b$, and because of symmetry conditions, the percentage of deformation obtained from expression (2) is the same for lattice constant $b$. Once the cell is generated with the elongated lattice constants according with expression 2, their values are kept unchanged and a relaxation of atomic positions takes place. The percentages of deformation ranges from 1% to 8% in steps of 1%.

Let us emphasize that the deformation by compressive strain has not been covered in this work for two main reasons: first, if a compressive strain is applied to a monolayer, it turns unstable as discussed later. Secondly, when systems are subjected to compressive strain, they can easily undergo rippling or crumpling leading to out-of-plane deformations, which makes the system unstable. In general, when compressive strain is applied to a 2D nanosystem, two types of instabilities may be induced: one related with fractures and another associated with buckling [60]. Fracture occurs when the load exceeds the limits of critical compressive stress of the material. On regards the buckling deformations, these are originated when the system is compressed and the atoms have less restriction to move up and down the plane in comparison with 3D systems until reaching the structure relaxation, this kind of deformations may induce dynamical instabilities in the systems [60]. These instabilities are mainly produced because of high electrostatic repulsions between atoms as a result of reduction of atomic distances. In the same way, compressive strains do not lead to the stabilization of phonon modes, i.e., phonon hardening instead of phonon softening if observed in systems under compression [53, 61, 62]. These instabilities are caused because of the decrease of the bond lengths as well as the occupancy of an antibonding orbitals [61], another reason is associated with the augmentation of buckling height which leads to an increase in the mixture of out of the plane and in the plane phonon modes [61].

In contrast, tensile strains have a positive impact in the 2D systems because, by this mechanism, it is possible to tune electronic properties (such as the electronic band structure, band gap and carrier mobilities), and several others, without losing stability, as long as the strains are moderate and do not exceed the breaking strain limits of the corresponding material. In general, 2D materials can withstand large tensile strains before rupture, for this reason, these materials are considered as promising candidates for several applications, especially in devices where a high performance must be warranted despite distorting, bending, or stretching deformations. Finally, the tensile strain originates the softening of some phonon modes leading to an improvement of the dynamical stability of 2D systems [53, 61, 62].

Table 1 summarizes the structural properties of the unstrained system together with the ones under tensile strain. The lattice constant of the unstrained system is in good agreement with other reports (see bottom of Table 1 for details).

**Table 1:** Lattice constants, bond lengths and bond angles of monolayers under tensile strain. Results from other work for unstrained system are added in the bottom of the table for comparison purposes.

| Percentage of deformation | a = b (Å) | B – P Bond length (Å) | B – P Bond angle (°) |
|---|---|---|---|
| 0% unstrained | 3.212* | 1.854† | 120 |
| 1% | 3.244 | 1.873 | 120 |
| 2% | 3.276 | 1.891 | 120 |
| 3% | 3.309 | 1.910 | 120 |
| 4% | 3.341 | 1.929 | 120 |
| 5% | 3.373 | 1.947 | 120 |
| 6% | 3.405 | 1.966 | 120 |
| 7% | 3.437 | 1.984 | 120 |
| 8% | 3.469 | 2.003 | 120 |
| other works (Å): * 3.18 [27, 63], 3.21 [28, 29, 30, 42, 50] | | | |
| † 1.83 [27, 63], 1.85 [28, 29, 30, 42, 50] | | | |

From Table 1 we can conclude that the bond lengths grow in the same proportion than the lattice constants. The bond angles remain constant, and the atoms coordinates are unaffected by the tensile strain.

### 3.2 Electronic properties

As a first step of the analysis, we describe the electronic properties of the unstrained system. Our results show that the monolayer behaves as a direct band-gap semiconductor. The valence band maximum (VBM) and the conduction band minimum (CBM) are both located at the K point (1/3, 1/3, 0) of the Brillouin zone (figure 2a). Projected density of states calculations reveal that the main contribution to the states around the Fermi level comes from $p_z$ orbitals of both, B and P atoms (figure 2b). The edge of the last valence band is formed by P-$p_z$ orbitals and a very low contribution of B-$p_z$ orbitals. In contrast, the edge of the first conduction band is mainly formed by B-$p_z$ with a very small contribution of P-$p_z$ orbitals. An important aspect to highlight is that in a wide range of energies below and above the Fermi level, there is a strong hybridization of $p_z$ orbitals from B and P atoms, which leads to a strong interaction of charge distributions.

In figure 2c, the total 2D charge distribution is presented. A top view (parallel to (0001) plane) shows that charge accumulates in the region between P and B atoms. In contrast, zones of charge depletion are observed around the center of the hexagon of the honeycomb pattern. In figure 2d, results of partial charge densities (or band decomposed charge densities) are presented. This corresponds to band-projected charge density, i.e., the contribution to total charge density coming from each single band. In our case, we plot the partial charge densities of the highest occupied band (HOB – last valence band) as well as the lowest unoccupied band (LUB – first conduction band).

A side view parallel to the $(\bar{2}110)$ plane shows the contribution to the charge density from the $p_z$ orbitals of B and P, which are oriented perpendicular to the *xy* plane. As previously mentioned, the main contribution to the last valence band and first conduction band comes from $p_z$ orbitals. Therefore, the charge distribution of figure 2d is in agreement with the projected density of states of figure 2b. The partial charge from HOB mainly accumulates around the P atom due to the high contribution of P-$p_z$ to the last valence band. In the same way, the partial charge from LUB is more localized around the B atom, consistent with the projected density of states which reveals that the main contribution to the first conduction band is due to B-$p_z$ orbitals.

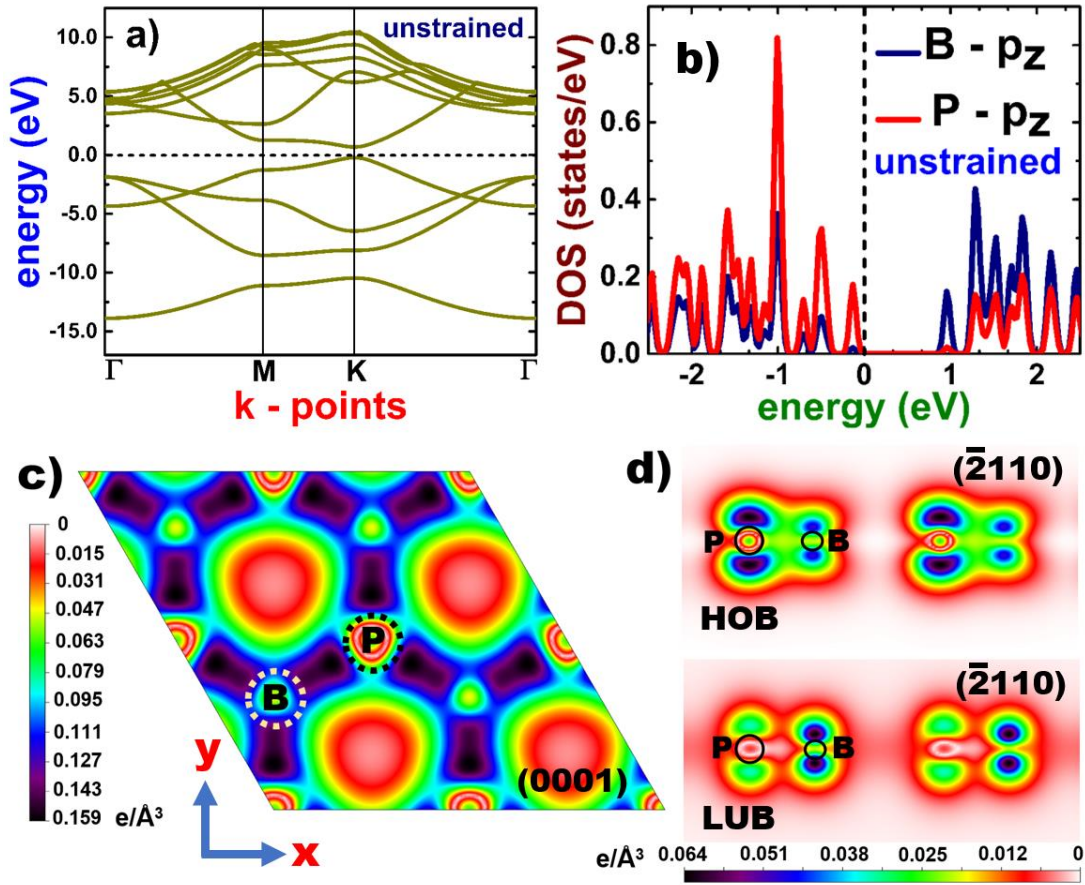

**Figure 2.** Electronic properties of the unstrained system. a) electronic band structure, b) projected density of states, c) 2D total charge density in the (0001) plane, and d) partial charge densities from the highest occupied band (HOB) and the lowest unoccupied band (LUB) in the $(\bar{2}110)$ plane.

Using the GGA approach, the computed band gap is equal to 0.857 eV. The underestimation of band gap obtained from standard DFT was corrected by using the $G_0W_0$ approach, from which, the corresponding computed value of the band gap is 1.832 eV.

The computed value within the GGA approximation is in good agreement with other reports, while our corrected value is in agreement with previous calculations based on the GW approach. Note that calculations using hybrid functionals underestimate the GW band gap. Let us now focus on the electronic properties of systems under tensile strain. Our results reveal that the electronic behavior is not affected by tensile strain, i.e., each system behaves as a direct band-gap semiconductor, with VBM and CBM both located at the K point of the Brillouin zone. There is no significant change in the band structures of the systems under deformation, especially in the region around the K point. In fact, the last valence band remains almost unchanged around the K point and there is only a small shift of the first conduction band to higher energies which opens the band gap (see figure 3 for details).

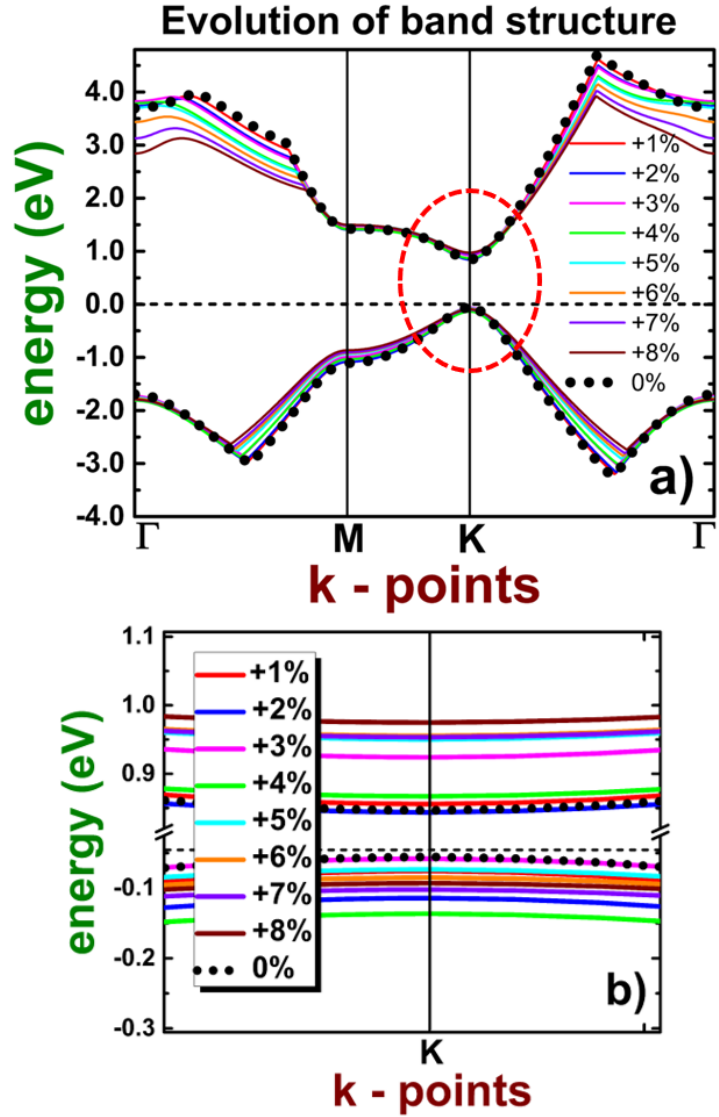

**Figure 3.** Electronic band structure as a function of tensile strain. Dotted line refers to the unstrained system. The energy reference corresponds to the Femi level, a) last valence band and first conduction band in the whole path of 1$^{st}$ Brillouin zone, b) zoomed view around the Fermi level in the energy range of last valence band maximum and first conduction band minimum.

The projected density of states of each strained system was also computed. According with the results, the strong hybridization of $p_z$ orbitals is kept in a wide range of energies, from −2.5 to 2.5 eV. For the states above the Fermi level, there is a uniform shift of all states (keeping the hybridization) to higher energies as the percentage of deformation increases, which opens the gap. For the states below the Fermi level, as the percentage of deformation increases, no significant changes occur in the density of states, for energies up to -0.3 eV (the region of the edge of the last valence band). However, for lower energies, from -0.3 to -2.5 eV we observe an accumulation of states (as the percentage of strain increases) coming from both, B and P atoms, and new peaks emerge. It is worth mentioning that the strong hybridization of $p_z$ orbitals remains despite the application of strain. Figure 4 depicts all-important aspects about the changes in the density of states of the systems under tensile strain.

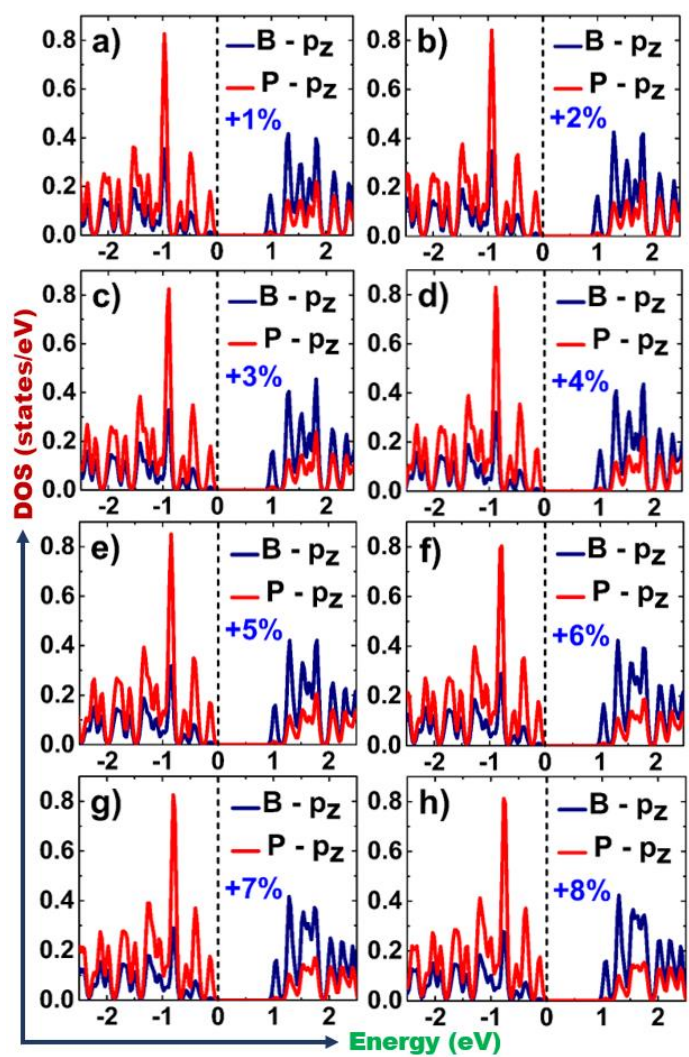

**Figure 4.** Projected density of states for systems under tensile strain. a) 1%, b) 2%, c) 3%, d) 4%, e) 5%, f) 6%, g) 7% and h) 8%.

We then computed the band gap for each system under strain within the $G_0W_0$ approach to correct the well-known underestimation of DFT calculations. Results are reported in Table 2 and in figure 5, which shows the behavior of the computed band gap using GGA and $G_0W_0$ approximations as function of percentage of deformation. The trend is observed to be the same in both approaches, the band gap increases linearly with a very small slope for percentages ranging from 1% to 5 %. For higher percentages (up to 8 %), the increase is slower, with a tendency to reach a constant value. We can conclude that, the sensitivity of band gap with respect tensile strain is in general slow and with a quasi-parabolic behavior. Results from both GGA and $G_0W_0$ were included for comparisons.

**Table 2:** Electronic band structure as a function of percentage of strain using GGA and $G_0W_0$ approaches.

| | Electronic band gap (eV) | |
|---|---|---|
| % of deformation | GGA | $G_0W_0$ |
| 0% (unstrained) | 0.898* | 1.833† |
| 1% | 0.927 | 1.874 |
| 2% | 0.954 | 1.910 |
| 3% | 0.978 | 1.935 |
| 4% | 1.000 | 1.959 |
| 5% | 1.019 | 1.983 |
| 6% | 1.037 | 2.001 |
| 7% | 1.052 | 2.017 |
| 8% | 1.065 | 2.029 |

other works (eV):
* 0.82 [27, 31], 0.84 [48], 0.88 [30], 0.90 [28], 0.91 [29].
† 1.74 [31], 1.81 [27].
HSE: 1.34 [29], 1.36 [25], 1.37 [26, 49], 1.39 [30, 48], 1.49 [44].
Tight-binding: 1.29 [45].

It is worth to mention that, we first tested the HSE06 hybrid functional to compute the electronic band-gap of unstrained system, the corresponding obtained value was 1.357 eV, which is in good agreement with previous reports [25, 26, 29, 30, 44, 48, 49], from Table 2 we can conclude that $G_0W_0$ provides a better approximation for the band-gap than hybrid functional does. For this reason, we used $G_0W_0$ to deal with band-gap calculations of strained systems. On the other hand, figures 6 and 7 depict a comparison among band structures (last valence band and first conduction band) obtained from standard DFT (GGA) and $G_0W_0$ approaches, the case considering the HSE06 hybrid functional was included for the unstrained system. As we can see from both figures, the results are qualitative similar, especially in the vicinity of high symmetry points Γ, M and K, but quantitative, the energies of band structures coming from $G_0W_0$ are shifted with respect GGA ones, which originates the band-gap opening. Let us recall that the k-points of band structures from $G_0W_0$ approach belong to special k-points in the 1st Brillouin zone defined by the Monkhorst-Pack scheme, since $G_0W_0$ calculations for quasiparticle energies are restricted to these points. We have considered the same points for the bands computed by using the HSE06 hybrid functional (for the unstrained system) in order to compare the $G_0W_0$ and HSE06 approaches.

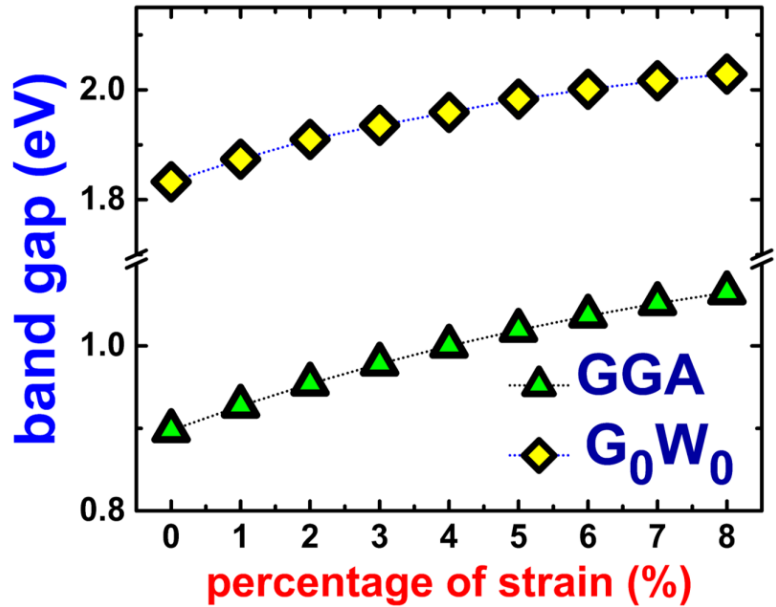

**Figure 5.** Electronic band gap computed within GGA and $G_0W_0$ approaches as a function of percentage of deformation.

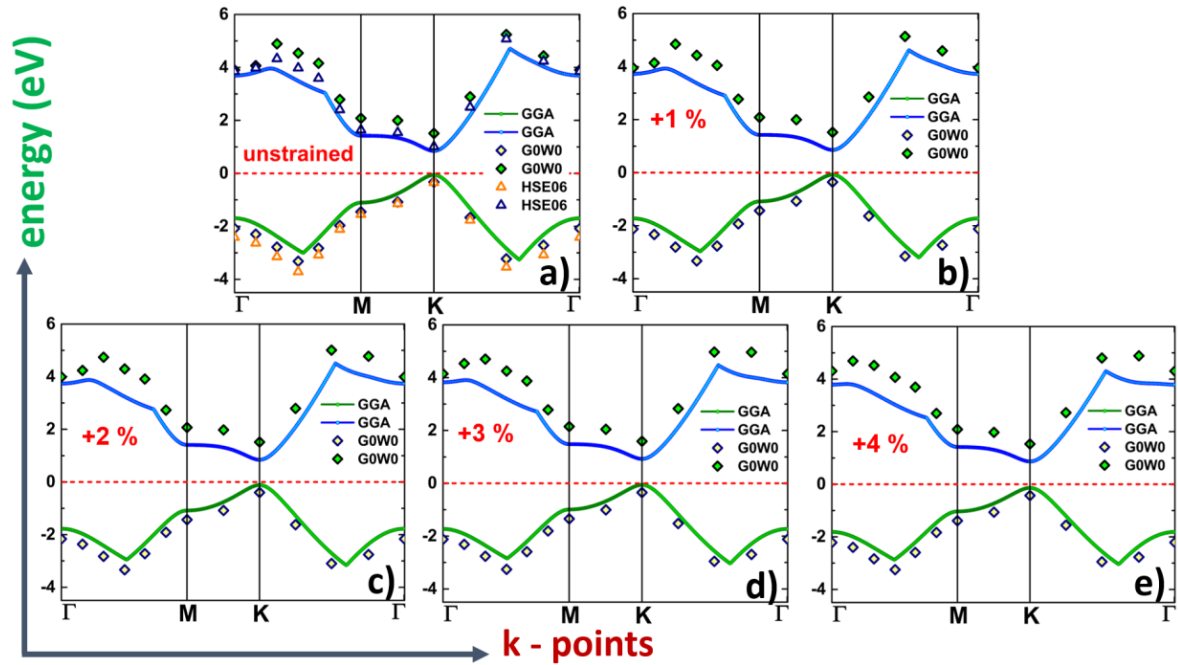

**Figure 6.** Last valence band and first conduction band computed using the GGA and $G_0W_0$ approaches for: a) unstrained system (including the bands computed using the HSE06 approach), b) +1 % strained, c) +2 % strained, d) +3 % strained, e) +4 % strained.

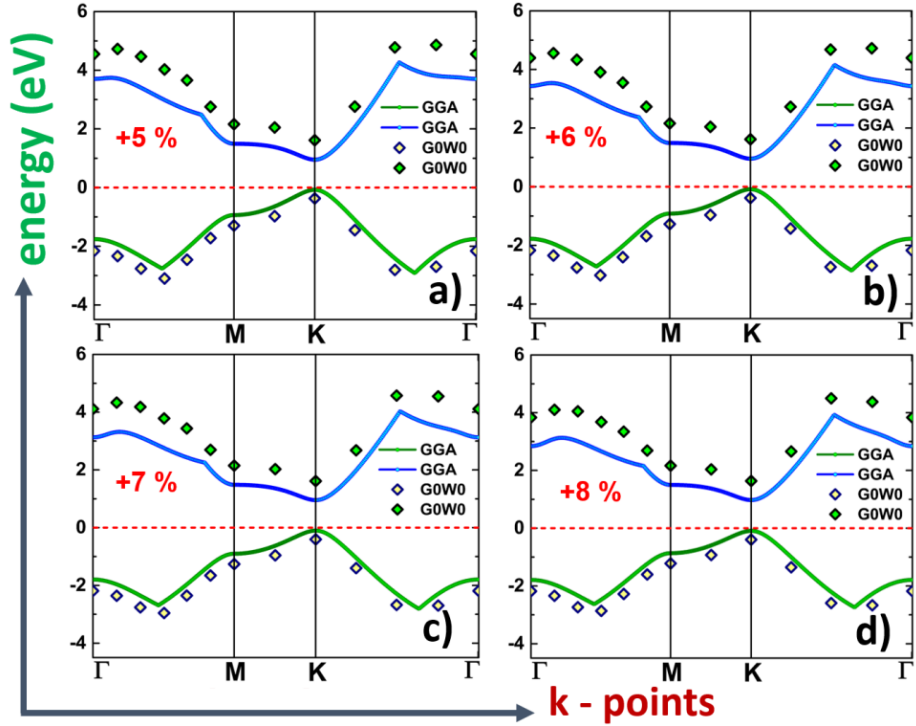

**Figure 7.** Last valence band and first conduction band computed using the GGA and $G_0W_0$ approaches for: a) +5 % strained, b) +6 % strained, c) +7 % strained, d) +8 % strained.

Let us now focus on explaining why the band gap increases with the deformation. The key point to understand this behavior is to describe what exactly happens with the states around the Fermi level. From figure 4 we observe that the hybridization of $p_z$ orbitals remains for all systems under tensile strain. In this way, the increment in band gap cannot be attributed to orbital overlap, or a contribution of other orbitals which is absent in the system without strain.

The change in atomic positions and bond lengths as a result of strain leads to a change in the charge distribution and the corresponding energies of the orbitals.

In this way, as a consequence of strain, it is expected that the Coulomb interaction between charges makes the bands to shift their energy levels (as shown in figure 3). This effect is also seen in the projected density of states where the orbitals move away from the Fermi level, which originates the change in the electronic band gap. Figure 8 shows the evolution of $p_z$ orbitals of B and P as a function of percentage of deformation.

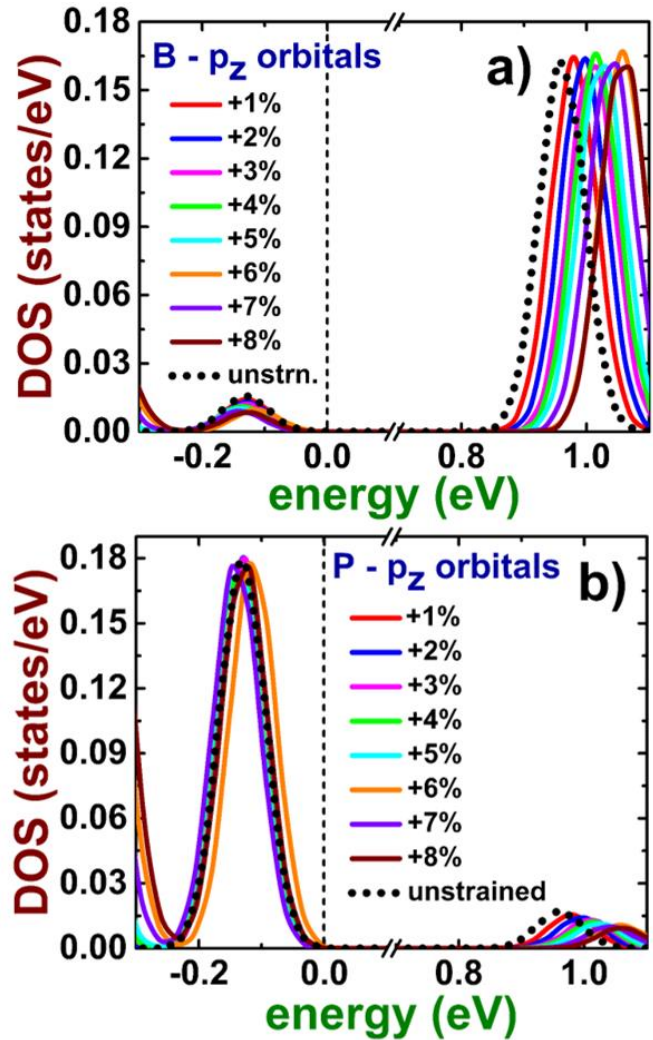

**Figure 8.** Projected density of states of $p_z$ orbitals as a function of percentage of deformation, a) B-atoms, b) P-atoms.

From figure 8, it is noted that the $p_z$ orbitals (of both atoms) below the Fermi level are almost unaffected by strain: only a small shift is observed in the orbitals. The change in the electronic band gap comes mainly from the shift of $p_z$ orbitals above the Fermi level, as percentage of deformation is increased. The orbitals move away to higher energies and this leads to band gap opening. It is worth keeping in mind that those $p_z$ orbitals of B and P shift simultaneously in order to retain the hybridization, as depicted in figure 4.

For a deep understanding of how the charge distribution is affected by strain, we have computed the charge density for the systems under strain. In figure 9, the total charge density in the (0001) plane is shown for representative systems. In the same way, the partial (band-decomposed) charge densities from both, HOB and LUB were computed. Results are shown in figure 10.

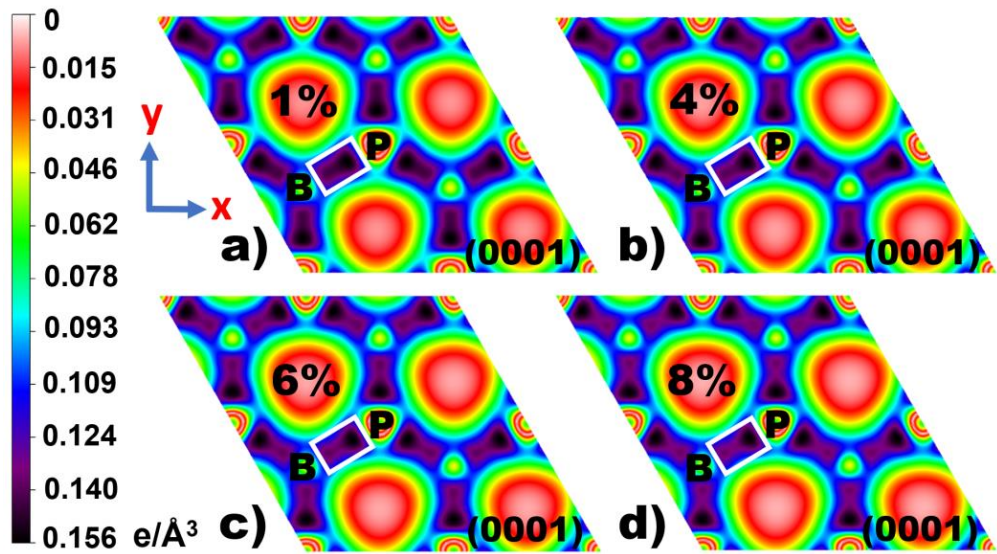

**Figure 9.** Total charge density in the (0001) plane for representative systems under strain. a) 1%, b) 4%, c) 6% and d) 8%.

As seen in figure 9, the charge accumulations are located along the bonds in-between B and P atoms (the region enclosed by the white rectangle), and the regions with less accumulation of charge are observed at the center of the hexagon formed by both atoms. As the percentage of strain is increased, the bond lengths elongate, and consequently, the charge accumulation along the bonds change. In systems with high values of strain, the charge mainly locates around the P atom, and a smaller accumulation of charge is observed around the B atom.

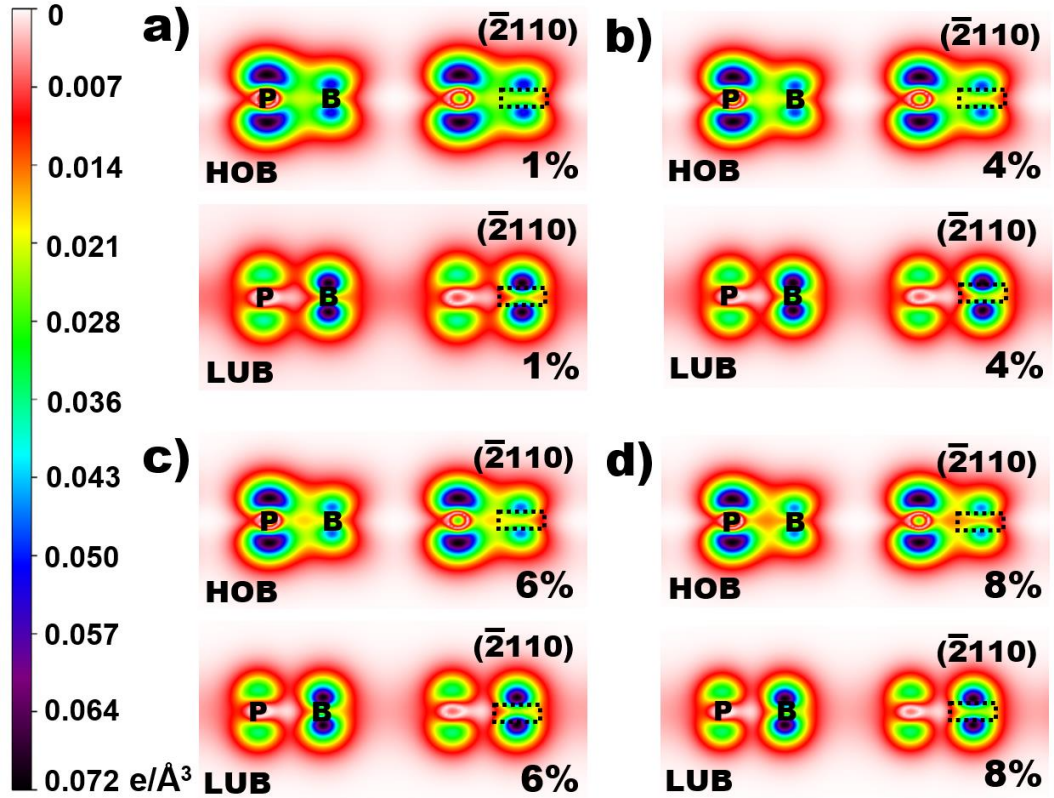

**Figure 10.** Partial charge distribution from highest occupied band (HOB) and lowest unoccupied band (LUB) in the ($\bar{2}$110) plane for representative systems under strain. a) 1%, b) 4%, c) 6% and d) 8%.

From figure 10, it is noted that the charge distributions (for both, HOB and LUB are mainly due to the $p_z$ orbitals) above and below the P atoms are not affected by strain. However, the charge distributions around the B atoms are affected by the tensile strain (see the region enclosed by a dashed rectangle). If we focus in the partial charge distributions of the HOB, we can see that, as the percentage of deformation increases, the charge accumulation above and below the B atoms decreases: the tensile strain produces charge redistribution around the B atoms. Besides, as the strain increases, a region with less accumulation of charge appears in the region between the B and P atoms. The opposite behavior is observed in the partial charge distributions of the LUB. In this case, as the tensile strain increases, a region of charge accumulation appears around the B atoms, while the charge density is unchanged in the region between B and P atoms. The effect of charge redistribution in this last case is lees notorious than the one observed in the LUB, in which, the charge distribution is more sensitive to strain. Although the tensile strain is applied in the *x-y* plane, the charge density is not only affected in the plane where the deformation takes place, but also in the region along the *z*-axis.

Finally, with regard to the trend observed in the band gap as a function of percentage of strain, the behavior can be explained in terms of how the charge distribution is affected by high strains. From the in-plane charge distribution of the unstrained system depicted in figure 2c, it is possible to notice that the charge is distributed practically evenly along the B-P bond. Once the system is strained, the atomic distance is increased, and if the deformation is small, the change in charge distribution is very sensitive due to the strong interaction between atomic orbitals, and there is a steep increase in the gap at small deformations (this can be seen in figure 9 where the charge accumulation is observed around P atom). However, for higher deformations, the interaction between atomic orbitals is less intense and no significant modification in charge distribution is observed, it remains mainly located around the P atom when strains are equal or greater than 5%, leading to a slower change in band gap. The same behavior occurs in charge distributions out of the plane coming from $p_z$-orbitals (figure 10). In this case, the charge is redistributed up and down the B atom, but, as observed for the in the plane case, the variation is more significant at low deformations than at higher ones. In summary, as the variations in charge distribution play the main role in the modification of electronic band gap, we can conclude that, the band gap is less sensitive to high deformations and changes slower because the charge density is not significantly affected when the strain in the system is high due to weaker interactions between atomic orbitals.

### 3.3 Computation of effective masses

In order to evaluate the mobilities of carriers (electrons and holes), the computation of effectives masses was performed. From the polynomial fitted curves of the band structures depicted in figure 11, we computed the effective masses of electrons and holes along two directions: K – Γ and K – M, since the VBM and the CBM are both located at point K, as shown in table 3.

For the unstrained system, we found that, in general, results are in good agreement with other reports. Some discrepancies emerge in the values of the effective masses of carriers along the K – M path. Results reveal smaller values than reported in reference [29], which indicate that the effective masses are of the same order of magnitude than the ones along the K – Γ path. However, this is in contradiction with the band structure shown in figure 2a (as higher curvature leads to lower effective mass), which clearly depicts that near the K point, the curvature in the K – M path is greater than the one along K – G for both carriers. It is expected that electron and holes are lighter along K – M than in the K – Γ direction. This is in agreement with our results listed in table 3.

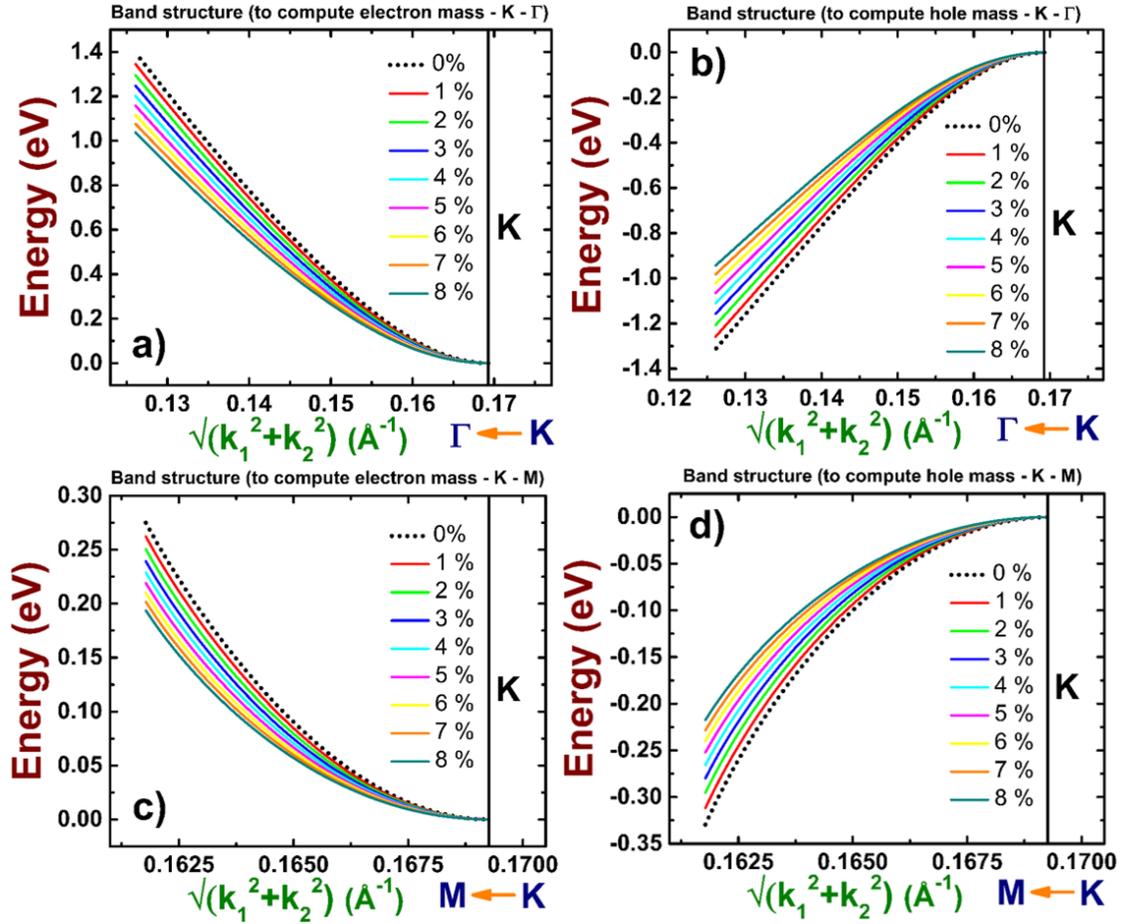

**Figure 11.** Segments of band structures along the symmetry lines of the Brillouin zone used to compute the effective masses: a) electron in the K → Γ direction, b) hole in the K → Γ direction, c) electron in the K → M direction, and d) hole in the K → Γ direction. $k_1$ and $k_2$ refers to the reduced coordinates of each *k*-point multiplied by the magnitude of vector $b_1$.

**Table 3.** Effective masses of carriers along the K – Γ, and K – M. Values from other reports are shown at the bottom for comparison purposes ($m_0$ refers the rest mass of electron).

| | Effective mass of carriers (in $m_0$ units) | | | |
|---|---|---|---|---|
| | K – Γ | | K – M | |
| % of deformation | $m_e$ | $m_h$ | $m_e$ | $m_h$ |
| 0% (unstrained)* | 0.1198* | -0.1183* | 0.0257† | -0.0232† |
| 1% | 0.1258 | -0.1246 | 0.0275 | -0.0249 |
| 2% | 0.1268 | -0.1262 | 0.0281 | -0.0256 |
| 3% | 0.1306 | -0.1304 | 0.0293 | -0.0268 |
| 4% | 0.1345 | -0.1348 | 0.0305 | -0.0280 |
| 5% | 0.1384 | -0.1393 | 0.0316 | -0.0292 |
| 6% | 0.1424 | -0.1437 | 0.0325 | -0.0303 |
| 7% | 0.1463 | -0.1487 | 0.0338 | -0.0316 |
| 8% | 0.1504 | -0.1535 | 0.0348 | -0.0329 |

Other works ($m_0$ units):
* $m_e$ = 0.120 [29], 0.190 [50], $m_h$ = -0.115 [29], -0.180 [50]
† $m_e$ = 0.151 [29], $m_h$ = -0.138 [29]

Figure 12 depicts the effective masses dependence on the tensile strain. The trend is the same for electrons and holes along both considered directions. The absolute value of effective masses increases in a quasilinear way as percentage of deformation augments. The values of effective masses remain in the same order of magnitude and the differences between them are quite small. In general, the effective masses are slightly sensitive to strain. The effective masses values in 2D-h-BP are smaller than the ones observed in other 2D systems such as phosphorene or transition metal dichalcogenides (TMD's). Therefore, it is expected to have very high mobility values of electrons and holes, especially in the K – M direction. This result shows the possibility of using 2D-h-BP in ultrafast electronic devices.

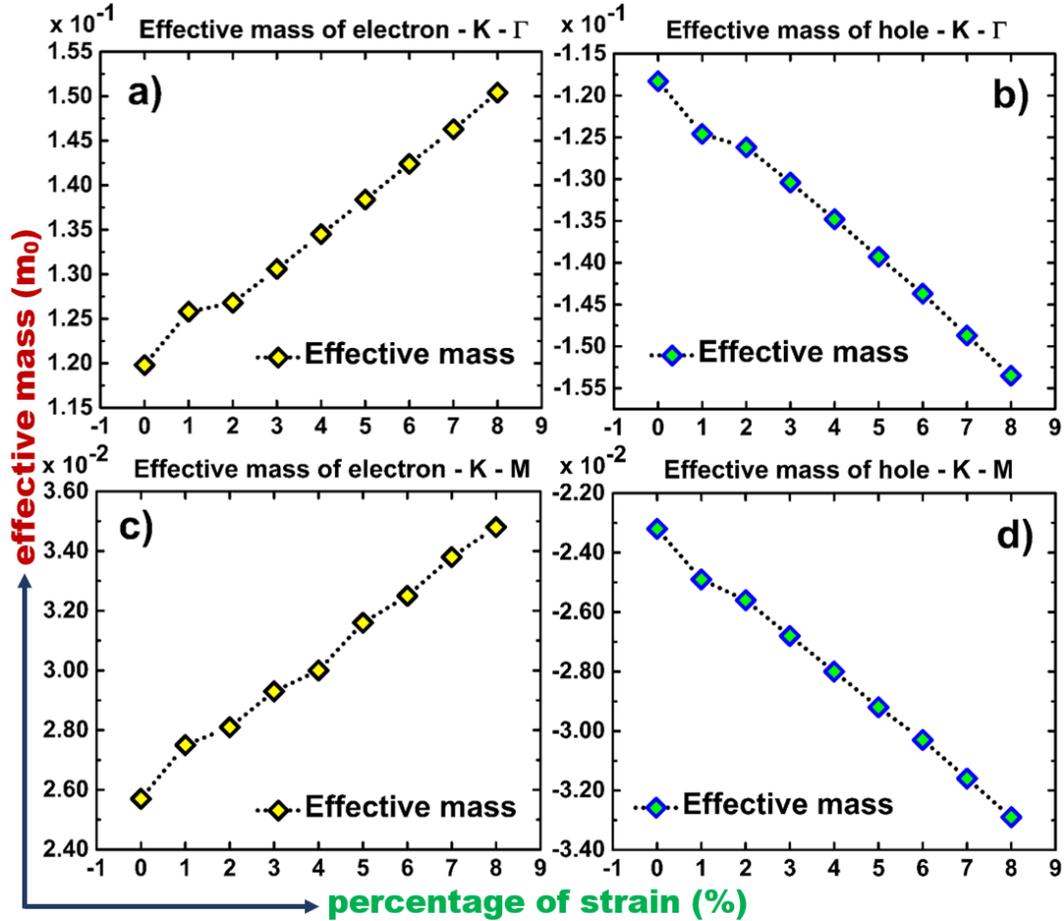

**Figure 12.** Effective masses of carriers as a function of percentages of deformation. a) effective mass of electron along K – Γ, b) effective mass od hole along K – Γ, c) effective mass of electron along K – M, d) effective mass of hole along K – M.

### 3.4 Analysis of stability

A deep study concerning the stability of systems under consideration is addressed in this section. First, we present the results of phonon spectra to assess the dynamical stability. In the same way, by the computation of elastic constants was possible to have an insight for exploring the mechanical stability. Finally, the calculation of cohesive energies allowed us to warranty the thermodynamical stability of all systems.

With regard to the dynamical stability analysis, we may say that it is a key point to assure the system construction in the laboratory. In this subsection, we describe the phonon dispersion for all systems under study. First of all, the unstrained system yields only positive frequencies (figure 13a), which indicates the system stability. In contrast, when compressing the system, negative frequencies appear, as a result of the increase in the interaction between B and P atoms and the tendency to break the $sp^2$ hybridization. Figure 13b shows clearly that the lowest acoustic branch presents a negative slope in the vicinity of the Gamma point, in the Gamma to M path. Therefore, this system is not stable under compression. Similar results are expected for larger compression, since the increase in the compression may drive the sp$^2$ hybridization break and the honeycomb-like structure loss.

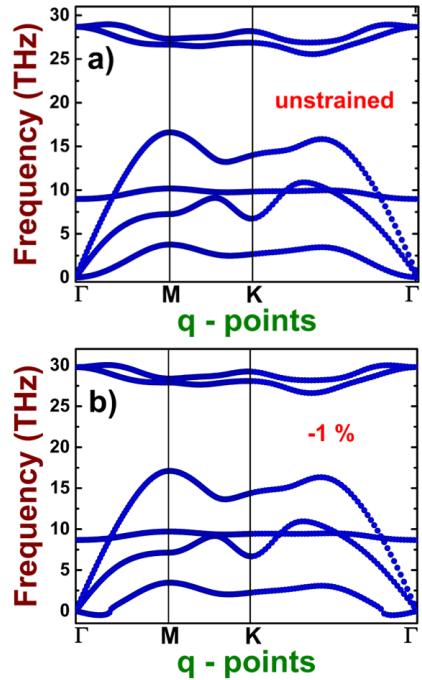

**Figure 13.** Phonon dispersion curves for a) unstrained monolayer, b) monolayer under compressive strain equal to -1%.

In contrast, when the 2D-h-BP system is under positive strain, the structure is stable until the 8% of isotropic in-plane deformation. Stability may be induced by the sp$^2$ hybridization enhancement since the sheet retains its planar honeycomb-like structure. This is confirmed by the non-negative frequencies in all phonon dispersion curves, see figure 14. In the unstrained system, the lowest acoustic branch shows a quasi-parabolic behavior near the Γ point. However, as a result of tensile strain, this behavior turns to linear, and the corresponding value of the slope becomes larger with the strain percentage increase (see figure 15 for details), indicating a stability enhancement. The evolution in the stability of this system when influenced by positive strain may assure its easy deposit in several substrates. Indeed, the substrate effect on the monolayer cannot be ruled out but if the substrate assures no breaking of the sp$^2$ like hybridization, the stability will remain. This assures its stability when including van der Waals heterostructures and substrates that assure weak interactions with the monolayer. In this kind of systems, the substrate effect may be more notorious in the electronic properties, modifying their energy gap by proximity to the substrate, as previously noted by Onam et al. [38].

The role of phonon gaps in the thermal conductivity of binary 2D sheets has been deeply discussed by Gu and Yang [64]. It is important to point out that phonon gaps appear in all our calculated phonon dispersions. In generally, phonon gaps typically emerge due to mass differences of species forming the compound [65] or the existence of different bond strengths (rigidity of the modes) [66]. In our case, when the 2D-h-BP monolayer is unstrained the phonon gap appears due to difference in masses, this is because we have the same kind of bonds all over the 2D sheet, so rigidity plays no role. However, when the system is under strain, the phonon gap evolves and becomes narrower as strain increases, evidencing that a change in bond strength is modulating the phonon gap. The described evidence points the importance of a study on the change of thermal conductivity in this material due to strain.

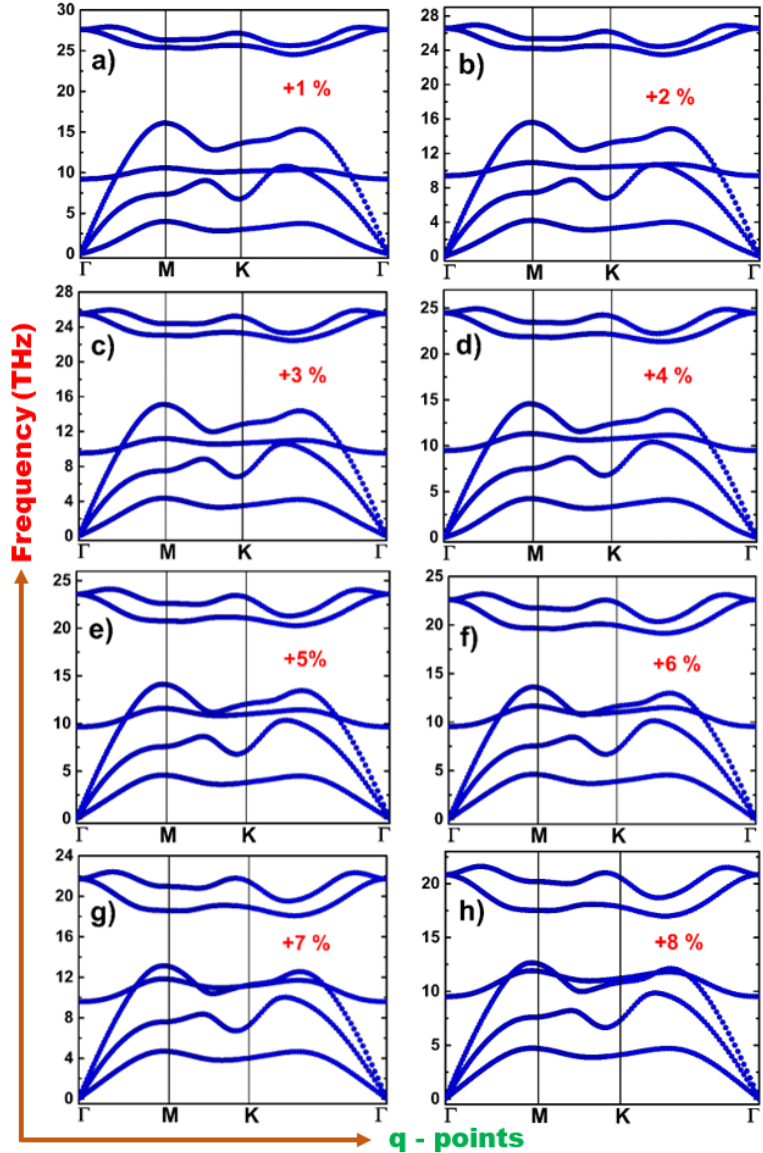

**Figure 14.** Phonon dispersion curves for systems under tensile strain with percentages of deformation equal to a) 1%, b) 2%, c) 3%, d) 4%, e) 5%, f) 6%, g) 7% and h) 8%.

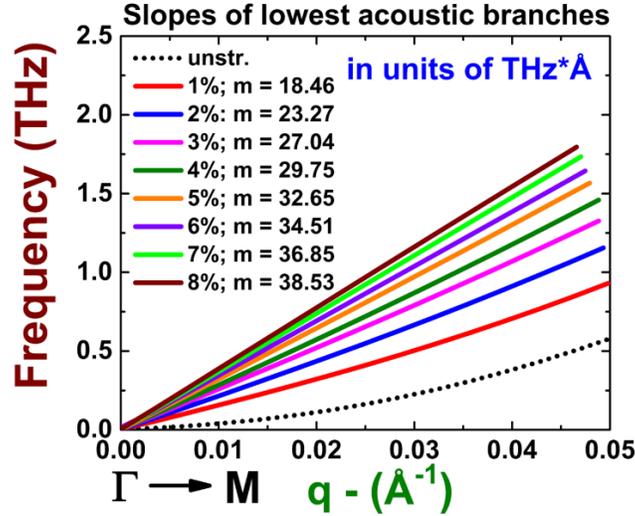

**Figure 15.** Segments of lowest acoustic branches in the Γ – M path for systems under tensile strain to compute the corresponding slope of each branch. The computation of slope for unstrained case has been omitted since it shows a quasi-parabolic behavior. Every segment of phonon dispersion has been fitted to a linear function for slopes calculations. Slopes values are given in units of THz · Å.

The mechanical stability can be established if the values of independent elastic constants satisfy certain necessary conditions. In our case, we have that 2D hexagonal boron phosphide belongs to the P6/mmm space group (No. 191), the same as graphene [67]. Therefore, 2D-hBP is considered as a graphene-like system. For this kind of structures, there are just two independent elastic constants [29, 30], namely $C_{11}$ and $C_{12}$. To warranty the mechanical stability of graphene-like systems, there are three necessary and sufficient conditions that must be satisfied [67-71]: $C_{11} > 0$, $C_{12} > 0$ and $C_{11} > C_{12}$. It is possible to obtain $C_{66}$ from $C_{11}$ and $C_{12}$ as: $2C_{66} = C_{11} - C_{12}$; as $C_{11} > C_{12}$ the condition $C_{66} > 0$ is also satisfied. According with our calculations, the three conditions are fulfilled not just for system free of strain, but also for all the other subjected to strain, the isotropy condition [29, 30, 67-71] for all cases is also fulfilled as $C_{11}$ is equal to $C_{22}$. The computed elastic constants are summarized in Table 4. Results concerning the system free of strain are in good agreement with previous reports.

In figure 16 it can be seen how the $C_{11}$ and $C_{12}$ elastic constants behave as a function of percentage of strain. In both cases, we can observe that the values of elastic constants decrease in a quasi-linear way, i.e., the system is less elastic as the deformation increases. Also, $C_{12}$ decreases to low values and tends to reach zero for strains higher than 8%, this result suggests that, for high values of tensile strain, $C_{12}$ could turn negative, leading to mechanical instability.

**Table 4.** Independent elastic constants (in N/m) of systems under study. In the bottom we have included the results for unstrained system of previous reports for comparison purposes.

| % of strain | $C_{11}$ | $C_{12}$ |
|---|---|---|
| unstrained | 144.780* | 39.391† |
| 1 | 137.828 | 34.484 |
| 2 | 130.369 | 29.562 |
| 3 | 122.971 | 25.270 |
| 4 | 114.716 | 20.904 |
| 5 | 108.902 | 17.742 |
| 6 | 101.298 | 14.429 |
| 7 | 94.626 | 11.738 |
| 8 | 88.470 | 9.501 |

* 145.900 [29], 146.285 [30].
† 38.800 [29], 38.753 [30].

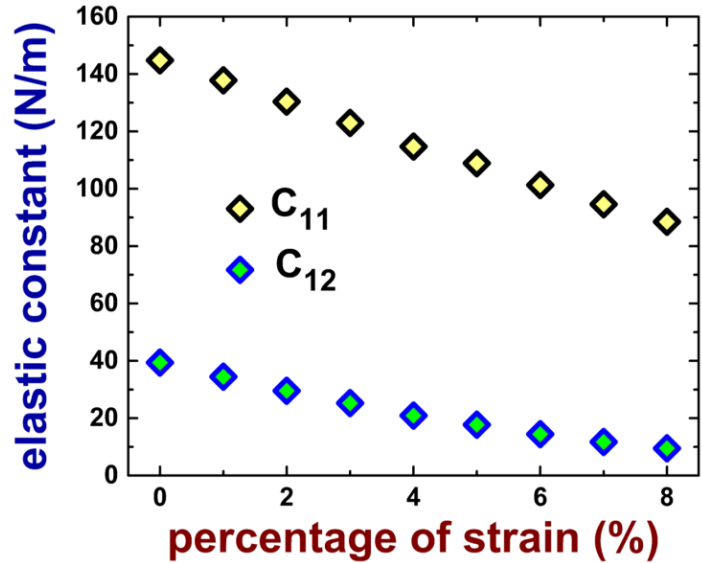

**Figure 16.** $C_{11}$ and $C_{12}$ independent elastic constants as a function of percentage of deformation.

Finally, to explore the thermodynamical stability, we have computed the cohesive energies using the following expression:

$$E_{coh} = E_{2D-hBP} - E_{B-isolated} - E_{P-isolated} \quad (3),$$

where:

$E_{coh}$ is the cohesive energy of 2D-hBP under consideration, $E_{2D-hBP}$ is the total energy of 2D-hBP under consideration, $E_{B-isolated}$ is the energy of the isolated B atom and $E_{P-isolated}$ is the energy of the isolated P atom.

The condition that is necessary to be fulfilled for warranting the thermodynamical stability of a system is that its corresponding value of cohesive energy must be negative, which means that the formation of the structure is favorable with respect to its constituent isolated atoms.

According with our calculation, this condition is fulfilled for all systems under study, which means that the 2D-hBP (with and without strain) is thermodynamically stable. Results for cohesive energies are presented in Table 5 and plotted in figure 17. The computed cohesive energy for unstrained system is in good agreement with other report (see table 5 for details). As we can see, the behavior is parabolic; cohesive energy is less negative as percentage of deformation increases (the most stable system is the unstrained one).

**Table 5.** Cohesive energies (in eV) for strained and unstrained 2D-hBP. The result from other repot for unstrained system is presented in the bottom of the table for comparison purposes.

| % of strain | Cohesive energy |
|---|---|
| unstrained | -9.860 † |
| 1 | -9.850 |
| 2 | -9.821 |
| 3 | -9.774 |
| 4 | -9.711 |
| 5 | -9.634 |
| 6 | -9.544 |
| 7 | -9.441 |
| 8 | -9.327 |
| † 9.50 (eV) [44] | |

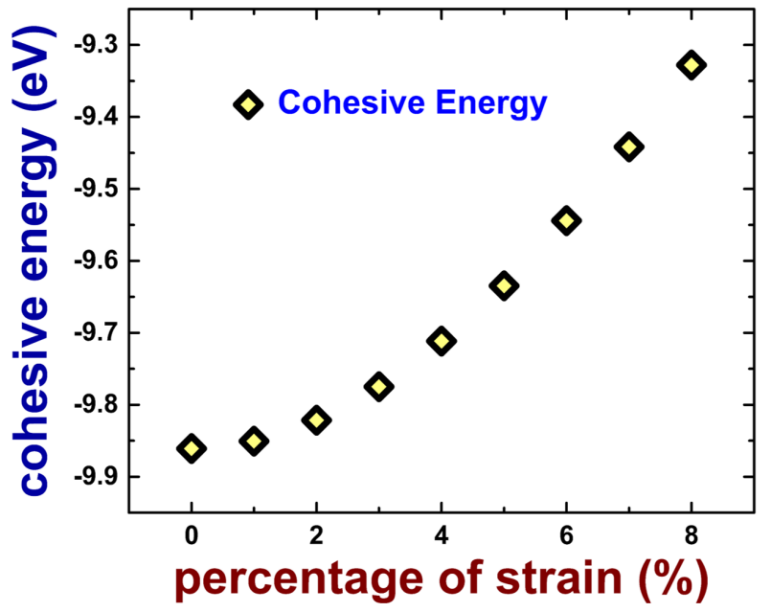

**Figure 17.** Results for cohesive energies of the systems under consideration.

As a final point and considering the results of our research, we can say that 2D-hBP is a promising candidate for several applications in flexible optoelectronics devices because of its high flexibility and strain sensitivity as a consequence of its small cross-sectional area (it is a one-atom thick flat structure) what warranties a low stiffness throughout the material.

2D-hBP can be applied in Field Effect Transistors (FETs) as the performance of these devices can be improved when the carrier mobilities are increased [53], which can be achieved in 2D-hBP when it is subjected to strain. On the other hand, when a tensile strain is applied in a 2D system, its electric resistance is changed because of the modification of the band-gap which is known as piezoresistive effect, this response is favorable for warranting a high performance in FETs [53], for this reason, 2D-hBP can be deposited in a flexible substrate for constructing a high-efficiency FET.

The piezoresistive effect observed in 2D-hBP (because of the modification of its band-gap by tensile strain) makes also possible its application in flexible strain sensors [53]. The gauge factor (GF) measures the resistance variation versus applied strain and it is an important parameter to determine if a flexible strain tensor will have a good performance, according with our results, 2D-hBP is expected to have a high GF, leading to a high sensitivity of flexible sensors based on this material.

Another potential application of 2D-hBP is in the fabrication of flexible photodetectors because of the combined effect of changing the lattice structure (the interaction area between 2D system and light is increased) together with the electronic band-gap [53]. The band-gap tuning makes possible to broaden the wavelengths of absorbed light in photodetectors, and consequently the photoresponsivity is also increased [52, 53]. In this way, by the application of tensile strain, the photocurrent of flexible photodetectors is expected to be high, which is attributed to the enhanced light absorption as a consequence of band-gap tuning. From our results, the values of band-gap in 2D-hBP when it is under tensile strain lie the region of the visible range making this material as good candidate for fabrication of flexible photodetectors.

## 4. Conclusions

A first principles study, based on the density functional theory, was performed to explore the electronic properties of hexagonal boron phosphide monolayers (2D-h-BP) subjected to tensile strain. The study was focused on the electronic band gap as a function of deformation percentage. The band gap was computed within the GW approach for correcting the underestimation from standard DFT. At low deformation percentages, in the range from 1% to 5% the band gap grows linearly with a low rate of change, for higher values of strain, up to 8%, the growth is slower and the band gap reaches saturation at a constant value of around 2 eV. The unstrained monolayer behaves as a direct band gap semiconductor and this behavior remains under deformation. In a wide range of energies, a strong $p_z$ orbitals hybridization coming from both B and P atoms is observed in the projected density of states, and is kept after deformation. In this way, the origin of the band gap opening is attributed to charge redistribution as a consequence of elongation of bond lengths. Because of this, the Coulomb interaction is less attractive when compared with the unstrained system, which induces a shift to higher energies in the $p_z$ orbitals above the Fermi level leading to the band gap opening. Electron and hole effective masses were computed in order to evaluate carrier mobility. Both carriers are able to move very fast along the two directions K – Γ and K – M of the first Brillouin zone, provided that effective masses are small, which makes possible the use of 2D-h-BP in ultrafast electronic devices. Finally, the phonon spectra of all systems were computed in order to assure dynamical stability. It was found that if the system is subjected to a compressive strain, it becomes unstable. However, the monolayer is able to resist deformations by tensile strain without losing dynamical stability. Finally, the computed elastic constants warrant the necessary conditions to assure mechanical stability, and the values of cohesive energies are all negative, indicating thermodynamical stability. All these results indicate that the formation of 2D-hBP structures is favorable.

## Acknowledgments

We thank DGAPA-UNAM projects IN101019, and IA100920, and CONACyT grant A1-S-9070, for partial financial support. Calculations were performed in the DGCTIC-UNAM Supercomputing Center and at LNSBUAP, projects LANCAD-UNAM-DGTIC-051, LANCAD-UNAM-DGTIC-368, and LANCAD-UNAM-DGTIC-382. J.M.G.H. acknowledges CONACyT, while R.P.P. and N.F.E. acknowledges DGAPA-UNAM for their postdoctoral positions at CNyN-UNAM. G.H.C. acknowledges the financial support of Cuerpo Académico de Física Computacional de la Materia Condensada (BUAP-CA-191). We also thank Aldo Rodriguez-Guerrero for their technical support.

## Data availability

The raw/processed data required to reproduce these findings cannot be shared at this time due to technical or time limitations.